# A stable 671 nm external cavity diode laser with output power exceeding 150 mW suitable for laser cooling of lithium atoms

SOURAV DUTTA[1,*] AND BUBAI RAHAMAN[1]

[1]*Tata Institute of Fundamental Research, 1 Homi Bhabha Road, Colaba, Mumbai 400005, India*
*Corresponding author: sourav.dutta@tifr.res.in*

**We report the design and performance of a Littrow-type 671 nm External Cavity Diode Laser (ECDL) that delivers output power greater than 150 mW and features enhanced passive stability. The main body of the ECDL is constructed using titanium to minimize temperature related frequency drifts. The laser diode is mounted in a cylindrical mount that allows vertical adjustments while maintaining thermal contact with the temperature stabilized base plate. The wavelength tuning is achieved by horizontal displacement of the diffraction grating about an optimal pivot point. The compact design increases the robustness and passive stability of the ECDL and the stiff but light-weight diffraction grating-arm reduces the susceptibility to low-frequency mechanical vibrations. The linewidth of the ECDL is ~360 kHz. We use the 671 nm ECDL, without any additional power amplification, for laser cooling and trapping of lithium atoms in a magneto-optical trap. This simple, low-cost ECDL design using off-the-shelf laser diodes without anti-reflection coating can also be adapted to other wavelengths.**

External Cavity Diode Lasers (ECDLs) are used extensively in many applications because of their narrow linewidth, frequency stability, frequency tunability, small foot print and low cost. ECDLs are the workhorse in a large number of atomic physics laboratories around the world and used extensively for laser cooling of atoms and ions because they satisfy the requirements of low linewidth, frequency stability and tunability [1–3]. In laser cooling experiments involving the D-lines of alkali atoms, an ECDL with the following characteristics is required. (a) The ECDL linewidth should be ≲1 MHz to address the atomic transition lines which typically have linewidths of 5-10 MHz. (b) For reliable day-to-day operation the frequency of the ECDL should remain stable within around 100 MHz for several hours without any active frequency stabilization. (c) It is desirable to have a mode-hop-free (MHF) frequency tunability of at least a GHz in order to visualize the Doppler-broadened atomic absorption spectrum while performing spectroscopy in a vapor cell. (d) In addition, for laser cooling experiments, an output power of the order 100 mW is typically required for efficient cooling and trapping of large number of atoms (>10$^8$) in a magneto-optical trap (MOT).

Laser diodes (LDs) are available in a wide range of wavelengths and can be used in an ECDL to fulfill these requirements for many atomic species. Several designs of ECDL of varied complexity exist in the literature and the choice of design is dictated by the specific requirements of the experiment. The Littrow design [1–4] is used in many experiments because it is simple, robust and delivers high output power. In some cases where the available output power is not of primary concern, a design [5,6] popularly known as the Littman-Metcalf design or designs based on interference-filters [7] are also used. Indeed for alkali atoms such as K, Rb and Cs, the output power of an ECDL is not of primary importance since the ECDL output can be amplified to several Watts of optical power using semiconductor tapered amplifiers (TAs) at 767 nm, 780 nm and 852 nm, respectively. However, due to technical difficulties with fabrication, such high power TAs are not available at 671 nm [8] – the wavelength required to address the $2S_{1/2}$-$2P_{1/2}$ and $2S_{1/2}$-$2P_{3/2}$ transitions of Li.

The technical challenges are also reflected in the fact that there are very few references describing home-built 671 nm ECDLs [3,9]. Even the commercially available ECDLs at 671 nm have a maximum power of ∼ 30 mW which can be amplified using TAs to a maximum output power of ∼ 400 mW. However, the beam profile of the TA output is non-Gaussian which leads to poor fiber coupling efficiencies and the TA chips are prone to degradation over time. Moreover, to the best of our knowledge, the sole source for these 671 nm TA chips is Toptica-Eagleyard Photonics and they have faced technical problems in producing these TA chips consistently and reliably [8]. Some research groups, including one of us, have previously implemented injection locking schemes [10] where the output of the ECDL is amplified using another laser diode [11] – however, in practice, the success has been limited because of the extreme sensitivity of injection locking to mechanical vibrations and temperature drifts. Some other research groups have resorted to all-solid-state frequency doubled laser systems [12] to overcome these problems but this is an extremely expensive alternative pushing the cost to $100k range. Given the scenario, it is imperative to invest in developing a stable, low cost, high power ECDL system that is suitable for Li laser cooling experiments without the need for a TA, injection locking or frequency doubling.

In the article, we report the design and performance of a Littrow-type 671 nm ECDL that delivers an output power >150 mW and features enhanced stability against frequency drifts. We demonstrate the application of the ECDL by implementing a MOT of $^7$Li atoms with the ECDL as the sole source i.e. without a TA. The main design ingenuities of our ECDL design are as follows. (a) We use off-the-shelf high power LDs, either Ushio HL65223DG or Ushio HL67001DG, with free-running wavelengths 660 nm and 675 nm, respectively. The lasing wavelength is tuned close to 671 nm by operating the LDs at 60°C and 16°C, respectively. Both the LDs have a maximum output power of ~220 mW in free-running condition, do not have any special anti-reflection (AR) coating and are readily available and inexpensive. They haven't previously been utilized for construction of ECDLs. (b) The main body of the ECDL is constructed using titanium (grade 5), as opposed to the more commonly used aluminium [13,14], because the coefficient of linear thermal expansion of titanium [~8.6 μm/(m·K)] is lower than aluminium [~23 μm/(m·K)] and the thermal conductivity of titanium [~6.7 W/(m·K)] is lower than aluminium [~237 W/(m·K)], although their volumetric heat capacities are similar [~2.4 J/(cm$^3$·K)]. The lower thermal expansion implies that the cavity length is less prone to drifts caused by temperature fluctuations, thereby making the ECDL frequency relatively more stable. The lower thermal conductivity is advantageous because the laser cavity is almost unaffected by sudden environmental temperature fluctuations. This helps in maintaining the single mode operation of the laser and eliminates sudden mode hops. (c) The MHF frequency tuning is achieved by horizontal displacement of the diffraction grating about an optimal pivot point [15]. This reduces the need to simultaneously tune the laser current [16] which can potentially lead to mode hops in high power non-AR-coated red LDs. The higher probability of mode-hops in high power non-AR-coated LDs is attributed to the longer length of the chip active area (required to accommodate higher optical power density) which results in internal cavity modes that are relatively closely spaced in frequency, enhancing the competition between adjacent modes.

*Design and Construction:* A schematic drawing of the ECDL is shown is Fig. 1. An off-the-shelf commercial laser current driver (Thorlabs LDC205C) is used to control the LD current. The LD is mounted in a collimation lens package (CLP) with a high numerical aperture of 0.55 (Thorlabs LT230P-B) and adjusted so that the laser beam profile is symmetrical at a distance of ~2 m. The CLP is then housed in the cylindrical LD holder that can be rotated for vertical alignment of the Littrow laser cavity. The laser holder and the diffraction grating holder rest on a baseplate whose temperature is controlled by a Peltier thermoelectric cooler (TEC) element. A commercial temperature controller (Thorlabs TED200C) in conjunction with a thermistor is used to maintain the temperature of the entire laser cavity. A holographic diffraction grating (Thorlabs GH13-24V), with 2400 lines/mm ($\equiv 1/d$) and 12% diffraction efficiency for the incident light with polarization parallel to the grooves, is used in Littrow configuration to feed the 1$^{st}$ order diffracted beam back into the laser diode. The diffraction grating holder is designed with a wedge so that the plane of diffraction grating intersects the pivot point about which the grating rotates (see Fig. 1). The pivot point is positioned at the intersection of the planes of the diffraction grating and the back face of the LD, shown as dotted lines in Fig. 1, to enable optimal frequency tuning characteristics. The course wavelength tuning (few nm) is achieved by manual displacement of the grating arm using a finely threaded adjustment screw, while the fine frequency tuning is achieved by a piezoelectric transducer (PZT) stack (Thorlabs AE0505D08F) placed behind the grating arm. The optimal location of the pivot point ensures that the two conditions, $n\lambda/2 = L$ and $2d\,sin\theta = \lambda$, for lasing in a single longitudinal mode of wavelength $\lambda$ remains satisfied as the cavity length $L$ ($\approx$ 20 mm) and the angle $\theta$ ($\approx 53.6°$) are tuned using the PZT. The central part of the ECDL (shown in Fig. 1) is covered with a plastic enclosure to minimize air currents and temperature fluctuations. The entire assembly is placed on the TEC mounted on a 1"-thick aluminium base and covered with a 1/8"-thick aluminium enclosure. The presence of two layers of enclosures, one of plastic and one of aluminium, resulted in visibly better stability of the ECDL and reproducible day-to-day operation.

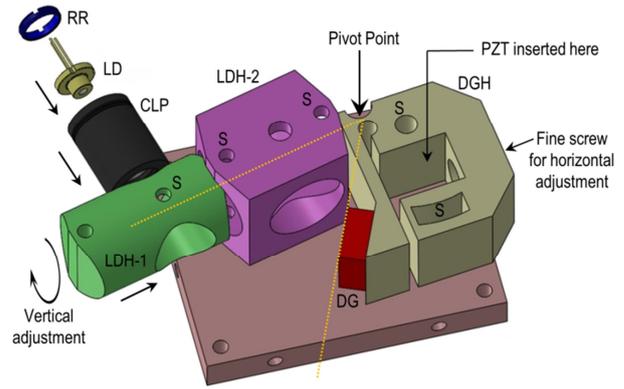

**Fig. 1.** Schematic drawing of the central part of the ECDL. The laser diode (LD) is held in the collimation lens package (CLP) with a retainer ring (RR). The CLP is inserted into the cylindrical laser diode holder (LDH-1), which is then inserted into LDH-2. LDH-1 can be rotated for vertical alignment of the ECDL and fixed into place with screws (S). The diffraction grating (DG) is glued to the diffraction grating holder (DGH) and placed such that the Pivot Point is at the interaction of the planes of the DG and the LD back, shown as dotted lines. A fine screw is used for manual horizontal adjustment of the grating arm and PZT is used to fine tune the wavelength of the ECDL. LDH-1, LDH-2, DGH and the base plate are made of titanium. This entire assembly sits on a TEC.

We constructed two ECDLs – in ECDL-1 we use the Ushio LD HL65223DG and in ECDL-2 we use the Ushio LD HL67001DG. We typically operate ECDL-1 (ECDL-2) at a laser diode current of ~230 mA (~180 mA), where the usable output optical power is ~ 100 mW (~115 mW). Figure 2 shows that optical power is excess of 150 mW can be achieved. The output optical power can be increased further by operating at higher currents (up to ~270 mA) but this was avoided to reduce the deleterious effects of mode competitions at higher powers and to prolong the lifetime of the LDs. Both ECDLs have similar characteristics and performance.

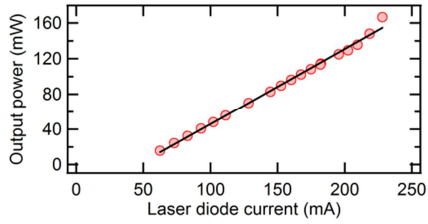

**Fig. 2.** Output optical power of ECDL-2 plotted against the laser diode current. The power was measured at currents that produced a ⁷Li saturated absorption spectrum. The spectrum is retained on changing the current by ∼ ± 0.5 mA around these values of the current, followed by a jump to a different wavelength and then returning back to the original wavelength at a different value of current.

*Characterization and Performance:* The ECDLs were designed and built with the goal of long term frequency stability. We found that the frequency drift of the ECDLs were < 50 MHz per hour and the Li spectrum was reproduced every day, after a warm-up time of around 20 minutes from turning it on, without the need for any realignment. Using a wavelength-meter of resolution 30 MHz, we measured that the ECDLs can be tuned by ∼ 4 GHz without mode-hops by applying a voltage ramp to the PZT alone (i.e. without any feed-forward to the laser diode current). With a current feed-forward, the MHF tuning range increased only slightly (6-10 GHz, depending on the laser current). This behavior is different from LDs with low power and/or with anti-reflection coating, where higher MHF tuning range can be achieved by tuning the injection current in conjunction with the PZT voltage [16]. We attribute the limited MHF tuning to the longer length of the chip and higher mode competition as discussed earlier. We note, however, that the tuning range is sufficient for absorption spectroscopy and laser cooling of atoms. We measured that the signal to total SSE ratio of the ECDL is comparable to that of commercial ECDLs.

To test the ECDLs we perform saturated absorption spectroscopy of Li vapor contained in a heat-pipe-oven (Fig. 3a). The heat-pipe-oven consists of a stainless steel tube (length 80 cm, diameter 4 cm) with two fused silica windows mounted using Viton O-rings at the two ends for optical access. Around five grams of solid lithium metal is placed in the center of the heat-pipe-oven and heated to around 325°C to generate Li vapor. The saturated absorption spectrum (SAS) for the ⁷Li $2S_{1/2} \to 2P_{3/2}$ transition is shown in Fig. 3b, where the features correspond to different ground and excited state hyperfine levels, $F$ and $F'$, respectively. The $2P_{3/2}$ state hyperfine levels are not resolved in the spectrum because the frequency difference between consecutive hyperfine levels, which lies in the range 3–9 MHz, are smaller than or of the same order as the natural linewidth (5.87 MHz) of the transition. In contrast, Fig. 3c shows the SAS for the ⁷Li $2S_{1/2} \to 2P_{1/2}$ transition along with the nearby ⁶Li $2S_{1/2} \to 2P_{3/2}$ transitions. The hyperfine levels of the ⁷Li $2P_{1/2}$ state are clearly resolved since the hyperfine level spacing (91.9 MHz) is larger than the natural linewidth and the laser linewidth. The hyperfine levels of the ⁶Li $2S_{1/2}$ state, spaced by 228.2 MHz, are also resolved. It is clear that the ECDL is suitable for high resolution spectroscopy.

For frequency stabilization, we locked the ECDLs to the $F = 2 \to F'$ transition of the SAS using the standard frequency modulation (current modulation) and feedback technique. With feedback to the PZT, we found that the ECDLs stays locked for hours in our typical lab environment (with pumps running, people talking, walking or working nearby). In order to estimate the linewidth of the laser, we measure the transmission spectrum through a scanning Fabry-Perot interferometer with a specified finesse ≥1500 and resolution ≤1 MHz. The measured width of the transmission peak is ≤2 MHz, which is a convolution of the width of the Fabry-Perot transmission and the ECDL linewidth. From this, we estimate the upper limit of the linewidth to be ≤1 MHz for both the ECDLs. For further investigation of the linewidth, we perform a heterodyne beat frequency measurement of the two ECDLs as follows. We lock ECDL-1 to the $F = 2 \to F'$ transition in the SAS with a low feedback bandwidth (<500 Hz) to stabilize it from frequency drifts due to environmental fluctuations (temperature, mechanical vibration etc.), while keeping the short-time (<1/500 s) linewidth unchanged. We do not lock the frequency of ECDL-2 (mainly because we do not have a separate heat-pipe-oven for spectroscopy) and tune its frequency close to ECDL-1 (within ∼100 MHz). We overlap the laser beams from ECDL-1 and ECDL-2 on a beam splitter and couple them into a single mode optical fiber for good spatial overlap. The output light is detected with an amplified photodiode (Thorlabs PDA10A2) of bandwidth 150 MHz and the signal is measured with a digital storage oscilloscope (DSO). The inbuilt FFT feature of the DSO is used to determine the beat signal (Fig. 4). We fit the signal to a Lorentzian function and obtain the combined linewidth of the two ECDLs to be ∼720 kHz. Assuming the two ECDLs to be identical but independent, the linewdith of each ECDL is ~360 kHz. The measured linewidth is comparable to ECDLs of other designs [3,14] and is sufficient for atomic spectroscopy and laser cooling experiments as demonstrated below. An important consideration for many experiments is the efficiency with which the laser beam can be coupled into a single mode polarization maintaining fiber that is required for transporting light and for obtaining a Gaussian transverse mode profile. We routinely achieve fiber coupling efficiencies of ~70 % for both the ECDLs using standard fiber couplers and optical fibers. The coupling efficiency is much higher than TAs, which have asymmetric transverse mode profiles with multiple lobes, and can be improved further with proper beam shaping arrangements.

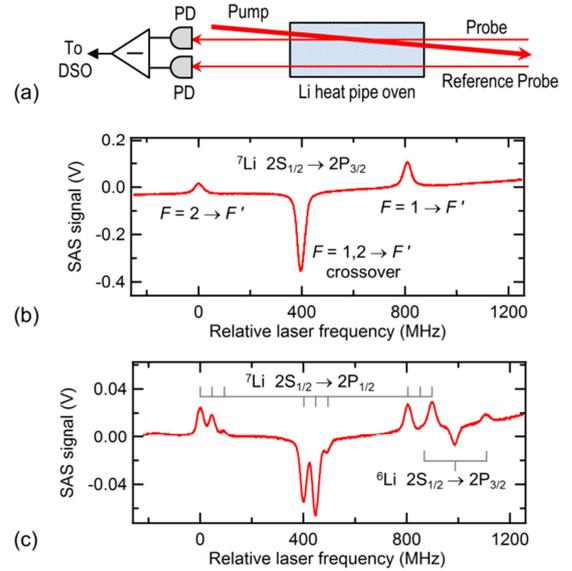

**Fig. 3.** (a) Schematic diagram of the SAS setup. The reference probe beam signal is used to subtract the Doppler-broadened absorption signal. (b) SAS of the $2S_{1/2}$ ($F$) $\to$ $2P_{3/2}$ ($F'$) transitions in ⁷Li. (c) SAS of the $2S_{1/2}$ ($F$) $\to$ $2P_{1/2}$ ($F'$) transitions in ⁷Li and the $2S_{1/2}$ ($F$) $\to$ $2P_{3/2}$ ($F'$) transitions in ⁶Li. The upright (inverted) tick marks indicate the various $F \to F'$ (crossover) transition. The laser frequency axis is not perfectly calibrated (it was generated by assuming that the laser frequency is linearly proportional to the voltage applied to the PZT). PD: Photodiode, DSO: Digital Storage Oscilloscope.

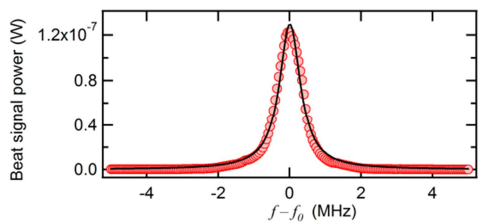

**Fig. 4.** Heterodyne beat frequency between ECDL-1 and ECDL-2 peaked at frequency $f_0$. The data (circles) was acquired on a DSO over a time interval of 2 μs and averaged over 10 sets. A Lorentzian fit (solid line) to the data gives a FWHM of ~720 kHz, which implies a laser linewidth of ~360 kHz.

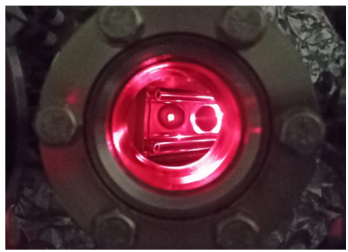

**Fig. 5.** A photograph of the MOT of $^7$Li atoms.

In order to demonstrate the use of the ECDL in laser cooling of Li atoms, we set up a MOT for $^7$Li with either ECDL-1 or ECDL-2 as the only source of light for the MOT and the Zeeman slower. The wavelengths required for the MOT and the decreasing-field Zeeman slower are obtained by appropriate frequency shifts of the laser beam using acousto-optic modulators (AOMs). An electro-optic modulator generates 803 MHz sidebands to address the $F = 1 \rightarrow F'$ repumping transition. The beams are then coupled into optical fibers. We get ~17 mW for the Zeeman slower beam and ~29 mW for the MOT beams which is then divided into three parts (these are the total powers including the carrier and the two sidebands). Each of the three beams have $1/e^2$ diameter of ~9 mm and are retro-reflected. The magnetic field gradient for the MOT is ~15 Gauss/cm. Additional details regarding the apparatus will be discussed elsewhere. In Fig. 5 we show a picture of the $^7$Li MOT taken with a cell-phone camera. We also collect the fluorescence using a set of lenses and detect the signal with a Si photodiode, from which we determine the atom number to be $1\times10^8$. This can be increased further by increasing the temperature of the lithium atomic beam source (currently operated at 350°C) and increasing the magnetic field of the Zeeman slower (currently the maximum field is 350 Gauss).

In summary, we report the design of a Littrow-type ECDL with enhanced passive stability and suitable for operation in a wide range of temperature. We demonstrate the operation of 671 nm ECDLs with output powers exceeding 150 mW and linewidth below 1 MHz that is suitable for high resolution spectroscopy and laser cooling of Li without the need for an optical amplifier. The improved stability of the ECDL originates from careful design and choice of titanium for the construction of the ECDL. The ECDLs were constructed using off-the-shelf LDs without any special AR coating which makes them attractive and low-cost alternatives to commercial systems. We have also tested the ECDL design with 852 nm LDs, to be used for laser cooling of $^{133}$Cs, and obtained similar results in terms of stability, linewidth, tunability etc. The ECDL design can be adopted for operation at other wavelengths.

**Funding.** Department of Atomic Energy, Government of India, Project Identification No. RTI4002.

**Acknowledgment**. We acknowledge the contributions of Kamal Kumar, Sagar Dam and Gorachand Das as project students during different stages of this work. Support from the TIFR Central Workshop is gratefully acknowledged.

**Disclosures.** The authors declare no conflicts of interest.

**Data availability.** Data and design underlying the results presented in this paper are not publicly available at this time but may be obtained from the authors upon reasonable request.

**REFERENCES**

[1] C. E. Wieman and L. Hollberg, Rev. Sci. Instrum. **62**, 1 (1991).
[2] K. B. MacAdam, A. Steinbach, and C. Wieman, Am. J. Phys. **60**, 1098 (1992).
[3] L. Ricci, M. Weidemuller, T. Esslinger, A. Hemmerich, C. Zimmermann, V. Vuletic, W. Konig, and T. W. Hsnsch, Opt. Comm. **117**, 541 (1995).
[4] T. W. Hansch, Appl. Opt. **11**, 895 (1972).
[5] I. Shoshan, N. N. Danon, and U. P. Oppenheim, J. Appl. Phys. **48**, 4495 (1977).
[6] M. G. Littman and H. J. Metcalf, Appl. Opt. **17**, 2224 (1978).
[7] X. Baillard, A. Gauguet, S. Bize, P. Lemonde, P. Laurent, A. Clairon, and P. Rosenbusch, Opt. Commun. **266**, 609 (2006).
[8] R. G. Hulet, J. H. V. Nguyen, and R. Senaratne, Rev. Sci. Instrum. **91**, 011101 (2020).
[9] K. G. Libbrecht, R. A. Boyd, P. A. Willems, T. L. Gustavson, and D. K. Kim, Am. J. Phys. **63**, 729 (1995).
[10] S. Dutta, Experimental Studies of LiRb: Spectroscopy and Ultracold Molecule Formation by Photoassociation, Ph.D. Thesis, Purdue University, 2013.
[11] A. C. Bordonalli, C. Walton, and A. J. Seeds, **17**, 328 (1999).
[12] U. Eismann, F. Gerbier, C. Canalias, A. Zukauskas, G. Trénec, J. Vigué, F. Chevy, and C. Salomon, Appl. Phys. B **106**, 25 (2012).
[13] E. C. Cook, P. J. Martin, T. L. Brown-Heft, J. C. Garman, and D. A. Steck, Rev. Sci. Instrum. **83**, 043101 (2012).
[14] L. Duca, E. Perego, F. Berto, and C. Sias, Opt. Lett. **46**, 2840 (2021).
[15] P. Mcnicholl and H. J. Metcalf, Appl. Opt. **24**, 2757 (1985).
[16] S. Dutta, D. S. Elliott, and Y. P. Chen, Appl. Phys. B **106**, 629 (2012).